\documentclass[prb,twocolumn,english,footinbib]{revtex4-2}
\usepackage{amsmath,amssymb,mathrsfs}
\usepackage{amsfonts,bm}
\usepackage{amsmath}
\usepackage{amssymb}
\usepackage{amsthm}
\usepackage{graphicx}
\usepackage{graphics}
\usepackage{subfigure}
\usepackage{color}
\usepackage{bm}
\usepackage{hyperref}
\usepackage{rotating}
\usepackage{comment}
\usepackage[colorinlistoftodos]{todonotes}
\usepackage{esint}
\usepackage{xfrac}
\usepackage{natbib}
\usepackage{upgreek}

\newcommand{\be}{\begin{equation} }
\newcommand{\ee}{\end{equation} }
\newcommand{\ba}{\begin{eqnarray} }
\newcommand{\ea}{\end{eqnarray} }

\newcommand{\bpm}{\begin{pmatrix}}
\newcommand{\epm}{\end{pmatrix}}
\newcommand{\bmm}{\begin{matrix}}
\newcommand{\emm}{\end{matrix}}

\newcommand{\la}{\label}
\newcommand{\p}{\partial}

\newcommand{\bea}{\begin{eqnarray}}
\newcommand{\eea}{\end{eqnarray}}

\makeatother

\usepackage{babel}

\begin{document}

%%%%%%%%%%%%%%%%%%%%%%%%

\title{Three Dimensional Odd Viscosity in Ferrofluids with Vorticity-Magnetization Coupling}

\author{Dylan Reynolds}
\author{Gustavo M. Monteiro}
\author{Sriram Ganeshan}

\affiliation{Department of Physics, City College, City University of New York, New York, NY 10031, USA }
\affiliation{CUNY Graduate Center, New York, NY 10031}

\date{\today}

%%%%%%%%%%%%%%%%%%%%%%%%

\begin{abstract}

Ferrofluids are a synthetic magnetic colloid consisting of magnetized nanoparticles surrounded by a repulsive surfactant layer. When subjected to an external magnetic field the ferrofluid acquires a macroscopic magnetization density which leads to magnetic behavior that is intricately coupled to the ambient fluid dynamics. Ferrofluids share several features with the chiral active fluids composed of unidirectionally spinning hematite cubes, which have been shown to possess a 2D non-dissipative odd viscosity term (Nature Physics,
15, 1188–1194 (2019)). In standard ferrofluid dynamics, 3D versions of parity breaking terms are not commonly observed, partly because of the small size of the magnetic particles. In this work,  we investigate if there are unique mechanisms in ferrofluids that can lead to a 3D odd viscosity term. Our results show that coupling the fluid vorticity ($\vec{\omega}$) to the magnetization ($\vec{M}$) with a term proportional to $\vec{\omega}\cdot\vec{M}$ leads to parity breaking terms in ferrofluid hydrodynamics, and results in a three dimensional odd viscosity term when the magnetization is relaxed to the direction of a uniform and static applied field. Hele-Shaw cells are commonly used devices to investigate ferrofluids and we demonstrate that this coupling reproduces the parity odd generalization of Darcy's Law discussed in a recent work (Phys. Rev. Fluids 7, 114201 (2022)). A potential experimental setup is discussed which may reveal the presence of this coupling in a ferrofluid confined to a Hele-Shaw cell.

\end{abstract}

%%%%%%%%%%%%%%%%%%%%%%%%

\maketitle

%%%%%%%%%%%%%%%%%%%%%%%%%%%%%%%%%%%%%%%%%%%%%%%%

\section{Introduction} \la{sec:intro}

Parity violation is ubiquitous in physical systems, from astronomical and geological phenomena at the largest of scales~\cite{yan2020hurricanes, jeong2019theOG, zemtsov2011rotation}, to the superfluid and quantum hall properties of electrons in condensed matter systems~\cite{avron1995viscosity,avron1998odd,tokatly2006magnetoelasticity,tokatly2007new,haldane2011geometrical,hoyos2012hall,bradlyn2012kubo,abanov2013effective, hoyos2014hall,laskin2015collective,can2015geometry,klevtsov2015geometric,scaffidi2017hydrodynamic,pellegrino2017nonlocal,berdyugin2019measuring,Monteiro2018nonresistivite,korving1966transverse}.   In describing the collective dynamics of many particle systems, parity breaking effects typically originate from the presence of intrinsic angular momentum at the level of the constituent particles. Recently much theoretical work has focused on describing so called ``active matter", systems in which energy is not conserved at the microscopic level. These systems naturally exhibit parity-breaking (parity odd) behavior~\cite{banerjee2017chiral, lubensky2021anal, monteiro2021hamiltonian}. Parity odd behavior in active matter systems can be described in a variety of ways. Odd elasticity is a framework that aims to capture these effects by studying the relation between stresses and strains for non-conservative microscopic interactions~\cite{scheibner2020elasticity}, while odd ideal gas descriptions study chiral collisions within a kinetic theory framework~\cite{fruchart2022oddgas}. Additionally, many chiral active systems are best described using a fluid description~\cite{banerjee2017odd, soni2018free}. 

In contrast to ordinary incompressible fluids, these active fluids manifest additional transport coefficients beyond shear viscosity. Perhaps the most famous of them is the rotational viscosity, which acts as a relaxation mechanism for the fluid particle angular momenta. In 2D, there is an extra viscosity known as ``odd" or ``Hall" viscosity, which preserves the fluid isotropy. Different from the other viscosity coefficients, odd viscosity is neither dissipative nor invariant under parity symmetry. Furthermore, its effects are only present for particular boundary conditions, such as free surface problems~\cite{ganeshan2017odd, abanov2018odd, abanov2018free}. On the other hand, parity breaking in 3D is incompatible with isotropy, leading to a much larger class of transport coefficients~\cite{vitelli2021long}. However, we recently showed that within a Hele-Shaw cell, the 3D parity breaking effects in incompressible fluids are revealed in a strikingly simple fashion~\cite{reynolds2021oddlaw,vitelli2021long}.

A ferrofluid is a type of 3D active matter fluid system that appears to possess several characteristics associated with parity-odd behavior. It is made up of magnetic nanoparticles coated in a hydrophobic surfactant layer and suspended in a carrier fluid. Each nanoparticle has an intrinsic magnetic moment, which is the sum of the magnetic moments generated by the atoms within it. Without an external magnetic field, thermal fluctuations prevent alignment of these moments, and the fluid is non-magnetic. When a substantial magnetic field is applied, the nanoparticles show collective behavior and the fluid becomes magnetized. For a detailed description of ferrofluids see Refs.~\onlinecite{rosensweig2013ferrohydrodynamics, rosenweig1987magneticfluids, neuringer1964ferrohydrodynamics, shliomis1974ferro, shliomis1988review, fang2022consistent, muller2002structure}. Two common laboratory setups are that of a steady applied field, where the magnetization density $\vec{M}$ aligns with the external field~\cite{rosensweig1983labyrinthine, vieu2018shape}, and a rotating applied field, in which $\vec{M}$ rotates with the external field with the same angular frequency but with a phase lag~\cite{rinaldi2002spin, chaves2008spin, shliomis2021rotate}.

The case of a rotating applied field, typically termed `spin-up' flow, in principle should result in parity odd behavior as the rotating field causes the magnetic particles to rotate, which introduces intrinsic angular momentum into the system at the microscopic level. However, in the limit of vanishing moment of inertia per particle, due to its small size ($I\sim r^2\rightarrow 0$), the microscopic angular momentum does not manifest itself at the hydrodynamic scale. Thus, there is no parity breaking behavior for ferrofluids in this setup. This should be contrasted to the configuration investigated in the recent chiral active colloids of Soni et al~\cite{soni2018free}, where the magnetic colloids, which are orders of magnitude larger than ferrofluid nanoparticles, shows experimental signatures of parity odd effects in the form of 2D odd viscosity. Furthermore, the case of steady applied field trivially does not lead to any parity odd transport coefficients, since no net angular momentum is generated at the macroscopic level.

Thus, it seems that in order for ferrofluids to manifest a 3D analog of odd viscosity, one needs the magnetization to play the role of the fluid intrinsic angular momentum, since the latter usually gets washed out due to the small size of the ferrofluid particles.  Following this intuition, in this paper we propose a coupling between magnetization density $\vec{M}$ and fluid vorticity $\vec{\omega}$, which after relaxing the direction of magnetization leads to the 3D odd viscosity in the case of a steady applied field. This coupling is motivated by the recent work of Markovich and Lubensky~\cite{lubensky2021anal}, where they have considered the intrinsic angular momentum coupling to fluid vorticity using a Poisson bracket formalism. The key feature of the magnetization-vorticity coupling is that the appearance of parity odd behavior at the macroscopic scales does not necessarily result from the angular momentum of the ferrofluid nanoparticles. In fact, this behavior should be present even in the limit of vanishing particle angular momentum.

The equations of motion for ferrohydrodynamics are usually derived from a detailed analysis of the microscopic forces and torques acting on the fluid~\cite{rosensweig2013ferrohydrodynamics}. However, to understand the origins of the magnetization-vorticity coupling it's more insightful to use a Poisson bracket (PB) formalism~\cite{sokolov2009hamiltonian, felderhof2011ferrobracket}, where the dissipative terms are incorporated through a dissipation function. Within this setup the relevant three dimensional parity breaking arises when we add the following {\it intrinsic} term to the ferrohydrodynamic Hamiltonian
\begin{align}
H_{M\omega} &=\frac{\gamma}{2} \int d^3 r \, \vec{\omega}\cdot \vec{M} , \label{eq:hamiltonian}
\end{align}
where $\vec{\omega}$ is the fluid vorticity, $\vec{M}$ is the magnetization per particle and $\gamma$ is a coupling constant that depends on the microscopic properties of the ferrofluid nanoparticles. As pointed out before, this term is analogous in form to that seen in Markovich and Lubensky~\cite{lubensky2021anal}, with the intrinsic angular momentum $\vec{\ell}$ replaced by $\gamma\,\vec{M}$. In that analysis it was seen that such term leads to what the authors referred to as 3D odd viscosity. This 3D odd viscosity breaks the fluid isotropy, hence, the corresponding viscosity term contains a transverse part (with respect to $\vec M$), which has the same form as the 2D odd viscosity, and a longitudinal part, which is only present in three dimensional systems.  

When this term is included in a ferrofluid system it can lead to a variety of parity-breaking behaviors. However, this paper will focus specifically on the case of a constant external field, and when the ferrofluid is confined to a Hele-Shaw cell. In this scenario the system is described by a modified version of Darcy's law, which was previously derived for general parity-odd flows in Ref.~\onlinecite{reynolds2021oddlaw}. Using a Hele-Shaw cell provides a reliable experimental setup to measure the strength of this new coupling.

This paper is organized as follows. In section \ref{sec:standard} we review the standard equations describing ferrofluids, and show how they can arise from a Hamiltonian and PB structure. In section \ref{sec:coupling} we introduce a vorticity magnetization term into the ferrofluid Hamiltonian and give the modified equations of motion. In section \ref{sec:heleshaw} we confine the ferrofluid to a Hele-Shaw cell and examine the modification to Darcy's Law.

\section{Review of Standard Ferrofluid Dynamics} \la{sec:standard}

In order to understand the significance of our modification, this section will provide an overview of the standard ferrofluid system and its governing equations. The ferrofluid nanoparticles are assumed to be evenly distributed throughout the carrier fluid, and the effects of the surfactant layer surrounding each particle are typically not considered. The ferrofluid is treated as a single entity with a constant density $\rho$, velocity $v_i$, pressure $P$, particle angular velocity $\Omega_i$, and magnetization per particle $M_i$. The applied field is denoted by $B_i$, and throughout this paper we will employ Einstein summation notation on repeated indices.

The fluid is assumed to be incompressible with good approximation, $\partial_i v_i=0$, and satisfy
\begin{align}
D_t v_i + \partial_i P &= \nu \nabla^2 v_i + M_j \partial_ j B_i - \Gamma\epsilon_{ijk}\partial_j\left(\omega_k - 2\Omega_k\right), \label{eq:govvel}
\end{align}
where the material derivative is defined as $D_t= \partial_t + v_j\partial_j$, $\epsilon_{ijk}$ is the Levi-Civita symbol,  $\nu$ is the kinematic shear viscosity, and $\Gamma$ is the kinematic rotational viscosity, which creates a drag force whenever the local particle rotation differs from (half) the local vorticity $\omega_i$. Note that in the above equation, and throughout this paper, the pressure $P$ has been scaled by the density, which has subsequently been set to unity. Conservation of local angular momentum gives the equation of motion for $\Omega_i$ to be
\begin{align}
I D_t \Omega_i  = \epsilon_{ijk} M_j B_k + 2\Gamma\left(\omega_i - 2\Omega_i \right), \label{eq:govomega}
\end{align}
where $I$ is the moment of inertia per particle, given by $I\sim r^2$, with $r$ being the typical radius of the ferrofluid particles. For the sake of simplicity, in our analysis we assume the nanoparticles are spherical, since the general case does not add much complexity. The governing equation for $M_i$ assumes that the magnetization is `frozen' in to the particle, and so the particles rotate as rigid objects,
\begin{align}
D_t M_i = \epsilon_{ijk}\Omega_j M_k . \label{eq:govmag}
\end{align}

A key assumption in ferrofluids is that the constituent particles are vanishingly small, and so we take the limit $I \rightarrow 0$. From Eq. (\ref{eq:govomega}) we see that this assumption implies that magnetic torque balances rotational friction
\begin{align}
2\Gamma\left(\omega_i - 2\Omega_i \right) = -\epsilon_{ijk} M_j B_k . \label{eq:relaxed}
\end{align}
We now solve Eq. (\ref{eq:relaxed}) for $\Omega_i$ and substitute into (\ref{eq:govvel}) and (\ref{eq:govmag}), which gives
\begin{align}
D_t v_i + \partial_i P &= \mu \nabla^2 v_i + M_j \partial_ j B_i +\frac{1}{2}\epsilon_{ijk}\partial_j\left(\epsilon_{klm}M_l B_m \right), \label{eq:standardv}\\
D_t M_i &= \frac{1}{2}\epsilon_{ijk}\omega_j M_k -\frac{1}{4\Gamma}\epsilon_{ijk}M_j\left(\epsilon_{klm}M_l B_m \right). \label{eq:standardmag}
\end{align}
Due to the smallness of the magnetic nanoparticles, thermal fluctuations can destroy the fluid magnetization for sufficiently low magnetic fields. A simplified model, appropriate for low magnetic field strengths, adds to Eq. (\ref{eq:standardmag}) a dissipation term of the form $-(M_i-M_i^0)/\tau$, where $\tau$ is a characteristic relaxation time. The magnitude of the equilibrium magnetization $M_i^0$ is given in terms of the Langevin function
\begin{align}
M^0_i=\frac{\mu B_i}{B}\left[\coth\left(\frac{\mu B}{k_B T}\right)-\frac{k_B T}{\mu B}\right], \label{eq:lang}
\end{align}
where $\mu$ is the magnetic moment of each particle. In the absence of an external applied field, or for high temperatures, the magnetic moments of each particle is `randomized' and the magnetization $M_i$ vanishes~\cite{chaves2008spin, felderhof2011ferrobracket}. A more detailed analysis takes the particle orientations to be random variables governed by Eq. (\ref{eq:govomega}), and averages over all angles. Taking the limit $I \rightarrow 0$ and substituting $\langle\Omega_i\rangle_T$ into the equation for $M_i$ yields a similar dissipation term, and modifies the rotational viscosity term~\cite{rosensweig2013ferrohydrodynamics, shliomis1988review}. For the purposes of this paper, either analysis will do, since the parity odd effects that we discuss are general and do not rely on specific microscopic details of the model.

The above equations of motion can also be derived starting from a Hamiltonian framework in combination with a dissipation function~\cite{sokolov2009hamiltonian, felderhof2011ferrobracket}. Since the ferrofluid is assumed to be incompressible, we will scale the Hamiltonian by $1/\rho$ ~\footnote{This is equivalent to $\rho=1$ everywhere.} and write the system in terms of the velocity $v_i$ and the angular momentum per particle $\ell_i=I\Omega_i$. Starting from a Hamiltonian of the form
\begin{align}
H = \int d^3 r \left[  \frac{1}{2}v_i v_i + \frac{1}{2I}\ell_i\ell_i   - M_i B_i  \right] \label{eq:ham}
\end{align}
and Poisson brackets give by
\begin{align}
\{v_i(\vec{r\,}),v_j(\vec{r}\,')\} &= \epsilon_{ijk}\,\omega_k(\vec{r\,})\,  \delta(\vec{r}-\vec{r}\,')\, , \label{eq:gwithg}\\
\{v_i(\vec{r\,}),\ell_j(\vec{r}\,')\} &=  \partial_i \ell_j(\vec{r\,})\,  \delta(\vec{r}-\vec{r}\,') \, ,\label{eq:gwithell}\\
\{v_i(\vec{r\,}),M_j(\vec{r}\,')\} &= \partial_i M_j(\vec{r\,})\,  \delta(\vec{r}-\vec{r}\,') \,, \label{eq:gwithm} \\
\{\ell_i(\vec{r\,}),\ell_j(\vec{r}\,')\} &= \epsilon_{ijk}\,\ell_k(\vec{r\,})\, \delta(\vec{r}-\vec{r}\,') \, ,\label{eq:ellwithell}\\
\{\ell_i(\vec{r\,}),M_j(\vec{r}\,')\} &= \epsilon_{ijk}\,M_k(\vec{r\,})\, \delta(\vec{r}-\vec{r}\,') \, \label{eq:ellwithm},
\end{align}
one can derive the equations of motion (\ref{eq:govvel}-\ref{eq:govmag}) using the following form of Hamilton's equations
\begin{align}
\partial_t v_i &= \{v_i, H\} - \frac{\delta R}{\delta v_i} - \partial_i \lambda , \label{eq:modbracket}\\
\partial_t \ell_i &= \{\ell_i, H\} - \frac{\delta R}{\delta \ell_i} ,\\
\partial_t M_i &= \{M_i, H\}- \frac{\delta R}{\delta M_i}, \label{eq:M}
\end{align}
where the effects of viscosity and dissipation are included by use of a dissipation function of the form
\begin{align}
R &=\int d^3 r\Big[ \frac{1}{2}\nu\left( \partial_i v_j + \partial_j v_i \right)^2  + \frac{1}{2}\Gamma\left(2\Omega_i - \epsilon_{ijk}\partial_j v_k \right)^2 \nonumber \\
&\quad\quad\quad\quad+ \frac{1}{2\tau}\left(M_i - M_i^{0} \right)^2 \Big] . \label{eq:r}
\end{align}
The  equation of motion (\ref{eq:modbracket}) is shifted by the gradient of a scalar function $\lambda$, which up to some redefinitions will be related to the fluid pressure. For incompressible fluids, the pressure does not follow from an equation of state, but works as a Lagrange multiplier to ensure that the velocity field dynamics is compatible with the incompressibility condition. Hence, the pressure function is completely determined by the flow and not an independent variable. Equations of motion~(\ref{eq:modbracket}-\ref{eq:M}) with the choice of Hamiltonian~(\ref{eq:ham}) and dissipation function~(\ref{eq:r}) lead to equations~(\ref{eq:govvel}-\ref{eq:govmag}) with the aforementioned relaxation term $-(M_i-M_i^0)/\tau$ added to the magnetization dynamics~(\ref{eq:govmag}).

%While providing a different means in arriving at the governing equations, the ferrofluid assumption of vanishing moment of inertia per particle $I$ is still required to accurately describe ferrofluids. The above ferrofluid equations of motion must be supplemented with appropriate boundary conditions depending on the specific problem of interest.

%, which can be stated as
%\begin{align}
%\textsf{Kinematic Condition} &\longrightarrow v_i \,\,\textsf{continuous} \label{eq:kinematic}\\
%\textsf{Stress Condition} &\longrightarrow T_{ij} \,\,\textsf{continuous} \label{eq:stress}\\
%\textsf{Normal Field} &\longrightarrow n_i B_i \,\,\textsf{continuous} \label{eq:normal}\\
%\textsf{Tangent Field} &\longrightarrow s_i H_i = J_i \label{eq:tangent},
%\end{align}
%where all quantities are evaluated at the boundary, be it a hard wall or free surface. Here, $\hat{n}$ and $\hat{s}$ are the normal and tangent unit vectors, respectively, to the surface, and $\vec{J}$ is the current density on the surface. For this work we take $\vec{J}=0$. 

 In the standard ferrofluid analysis, the assumption of vanishing moment of inertia per particle $I$ is imposed before setting $\tau\rightarrow 0$. In other words, the nanoparticle angular momenta relax much earlier than their magnetization. For uniform magnetic field, the resulting equation is simply the Navier-Stokes equation
\begin{equation}
D_tv_i=\nu\nabla^2v_i-\partial_iP\,,
\end{equation}
together with the incompressibility condition
\begin{equation}
    \partial_i v_i=0\,.
\end{equation}

\section{ Parity Breaking in Ferrofluids Via Vorticity-Magnetization Coupling} \la{sec:coupling}

The Hamiltonian and PB framework of ferrofluids provides a straightforward way to add modifications to the system. While many additional terms are in principle possible, not all terms are relevant or lead to interesting behavior. For example, one may add the term seen in Markovich et al~\cite{lubensky2021anal}
\begin{align}
H_{\ell\omega} &=  \int d^3 r\, \frac{1}{2}\ell_i\omega_i , \label{eq:lubterm}
\end{align}
which couples fluid vorticity to intrinsic angular momentum density. While this does lead parity breaking terms in the stress tensor, they are washed out under the ferrofluid assumption of vanishing moment of inertia. In our analysis we will keep this term to compare and make contact with our modification. As discussed earlier, this term is relevant in chiral active fluids of Soni et al, where the constituent particles are orders of magnitude larger than the conventional ferrofluids. One may also add quadratic terms of the form 
\begin{align}
H_{MM} = \int d^3 r\,\frac{1}{2}M_i M_i ,
\end{align}
however without modifying the brackets (\ref{eq:gwithg}) - (\ref{eq:ellwithm}), this term does not contribute to the equations of motion. If we wish to describe magnetic solids, or any motion of the magnetization relative to the particle orientation, we may add a nonvanishing bracket between $M_i$ and $M_j$ in the following way
\begin{align}
\{M_i(\vec{r\,}),M_j(\vec{r}\,')\} &= \frac{1}{\gamma}\epsilon_{ijk}\,M_k(\vec{r\,})\, \delta(\vec{r}-\vec{r}\,') \, .\label{eq:mwithm}
\end{align}
With the inclusion of this term, the equation of motion for $M_i$ becomes the Landau-Lifshitz-Gilbert equation, which would more precisely track the relaxation of the magnetization~\cite{lakshmanan2011fascinating}. In this work, we ignore the bracket ({\ref{eq:mwithm}) and assume the magnetization is `frozen' into the ferrofluid particles. Physically this represents the fact that when attempting to align with the external magnetic field, it is more favorable for the particle to rotate than for the magnetization to rotate independently.

The full Hamiltonian we will study, including both $H_{M\omega}$ given in (\ref{eq:hamiltonian}) and $H_{\ell\omega}$ given in (\ref{eq:lubterm}), is then
\begin{align}
H = \int d^3 r \left[ \frac{1}{2}v_i v_i + \frac{1}{2I}\ell_i\ell_i +\frac{1}{2}\ell_i\omega_i +\frac{\gamma}{2}\omega_i M_i - M_i B_i \right] \,. \label{eq:fullham}
\end{align}
For our analysis we leave the coupling constant $\gamma$ as a free parameter to be investigated experimentally or from some more microscopic description. However, a heuristic analysis of the structure of the ferrofluid nanoparticles, and the atoms within, shows $\gamma$ to be inversely proportional to the molecule Land\'e $g$-factor. 

%{\fb Within the atom there are three potential sources for the magnetic diople; the intrinsic magnetic moments of the nucleons and electrons coming from their spin, and the dipole generated by any orbital angular momentum of the electrons. The vorticity magnetization coupling term $\mathcal{H}_{I}$ only couples the fluid vorticity to the orbital angular momentum, and not the intrinsic spin. In this sense, the coupling is of the same type as $\mathcal{H}_{l}$, since both relate fluid vorticity to some angular momentum or rotation within the constituent particles.}

We now compute the governing equations resulting from the Hamiltonian (\ref{eq:fullham}), using the brackets (\ref{eq:gwithg}) - (\ref{eq:ellwithm}), along with the dissipation function (\ref{eq:r}). We obtain
\begin{align}
D_t v_i + \partial_i P = \nu \nabla^2 v_i &+ M_j \partial_ i B_j - \Gamma\epsilon_{ijk}\partial_j\left(\omega_k - 2\Omega_k\right)\nonumber \\
& + \frac{1}{2}\omega_j \partial_j \left( I\Omega_i + \gamma M_i \right) , \label{eq:newvel}
\end{align}
\begin{align}
I D_t \Omega_i  =& \epsilon_{ijk} M_j B_k + 2\Gamma\left(\omega_i - 2\Omega_i \right) \nonumber \\
&+ \frac{1}{2}\epsilon_{ijk}\omega_j \left( I\Omega_k + \gamma M_k \right) \nonumber \\
&-\frac{I}{2}\epsilon_{jkm}\partial_j\Omega_i \partial_k\left( I\Omega_m + \gamma M_m \right), \label{eq:newomega}
\end{align}
\begin{align}
D_t M_i &= \epsilon_{ijk}\Omega_j M_k + \frac{1}{2}\epsilon_{ijk}\omega_j M_k -\frac{1}{\tau}(M_i-M_i^0)  \nonumber \\
&\quad\quad -\frac{1}{2}\epsilon_{jkm}\partial_j M_i \partial_k\left( I\Omega_m + \gamma M_m \right)   . \label{eq:newmag}
\end{align}
where $P=\lambda-\frac{1}{2}v_iv_i-\frac{I}{2}\Omega_i\Omega_i+M_i B_i $ is the modified pressure. The terms involving $I\Omega_i + \gamma M_i$ stem from $H_{\ell\omega}$ and $H_{M\omega}$, while the second term on the right hand side of (\ref{eq:newmag}) comes from $H_{\ell\omega}$ alone. The set of equations (\ref{eq:newvel}) - (\ref{eq:newmag}) model a wide class of magnetic fluids, of any particle size and any orientation of the magnetic field and magnetization.

%Under the right conditions they describe the setup in Soni et al~\cite{soni2018free}, with the additional terms in (\ref{eq:newvel}) being the source of the odd viscosity seen there~\cite{lubensky2021anal} {\fb Soni et al yields 2D odd viscosity and we have a 3D hydro, expand the discussion to include how we can obtain an eddective 2D hydro}.

Upon taking the ferrofluid limit ($I \rightarrow 0$), Eq. (\ref{eq:newomega}) gives the modified torque balance equation
\begin{align}
2\Gamma\left(\omega_i - 2\Omega_i \right) + \epsilon_{ijk} M_j B_k + \frac{\gamma}{2}\epsilon_{ijk}\omega_j M_k = 0 \,, \label{eq:newbalance}
\end{align}
which shows that in the limit of vanishing particle size rotational friction is balanced not only by the magnetic torque, but also the torque coming from the misalignment of fluid vorticity and magnetization. Upon solving (\ref{eq:newbalance}) for $\Omega_i$ and substituting into the rest of the equations of motion, we get the final form of the modified ferrofluid equations
\begin{align}
D_t v_i +\partial_i P  &= M_j \partial_i B_j + \nu \nabla^2 v_i + \frac{\gamma}{2}\omega_j \partial_j  M_i \nonumber \\  
&\quad + \epsilon_{ijk}\partial_j \left(\frac{1}{2}\epsilon_{klm}M_l B_m + \frac{\gamma}{4}\epsilon_{klm}\omega_l M_m \right) , \label{eq:finalvel}\\
D_t M_i &= \epsilon_{ijk}\omega_j M_k - \frac{1}{4\Gamma}\epsilon_{ijk}M_j \left(\epsilon_{klm}M_l B_m\right) \nonumber \\
& \quad + \frac{\gamma}{8\Gamma}\epsilon_{ijk}M_j\left(\epsilon_{klm}M_l \omega_m\right) - \frac{1}{\tau} \left(M_i - M_i^0 \right)\nonumber \\
&\quad\quad -\frac{\gamma}{2}\epsilon_{jkm}\partial_j M_i \partial_k  M_m  \label{eq:finalmag}
\end{align}

In the case of a uniform magnetic field applied in the positive $z$ direction, the small magnetic relaxation time $\tau$ sets $M_i = |\vec{M}^0|  \delta_{iz}$, with $\vec{M}^0$ given by (\ref{eq:lang}). Equation (\ref{eq:finalvel}) can then be written as
\begin{align}
D_t v_i + \partial_i P &= \nu \nabla^2 v_i + \frac{\gamma}{4}M^0\partial_z \omega_i   . \label{eq:simplifiedv}
\end{align}
This last term breaks parity and is similar to the term obtained by Markovich and Lubensky after taking the incompressible limit, albeit with $M_z$ replacing the intrinsic angular momentum term $\ell_z=I \Omega_z$.

\section{Ferrofluids in Hele-Shaw cell: Experimental proposal} \la{sec:heleshaw}

%%%%%%%%%%%%%%
\begin{figure*}
\centering
\includegraphics[scale=0.45]{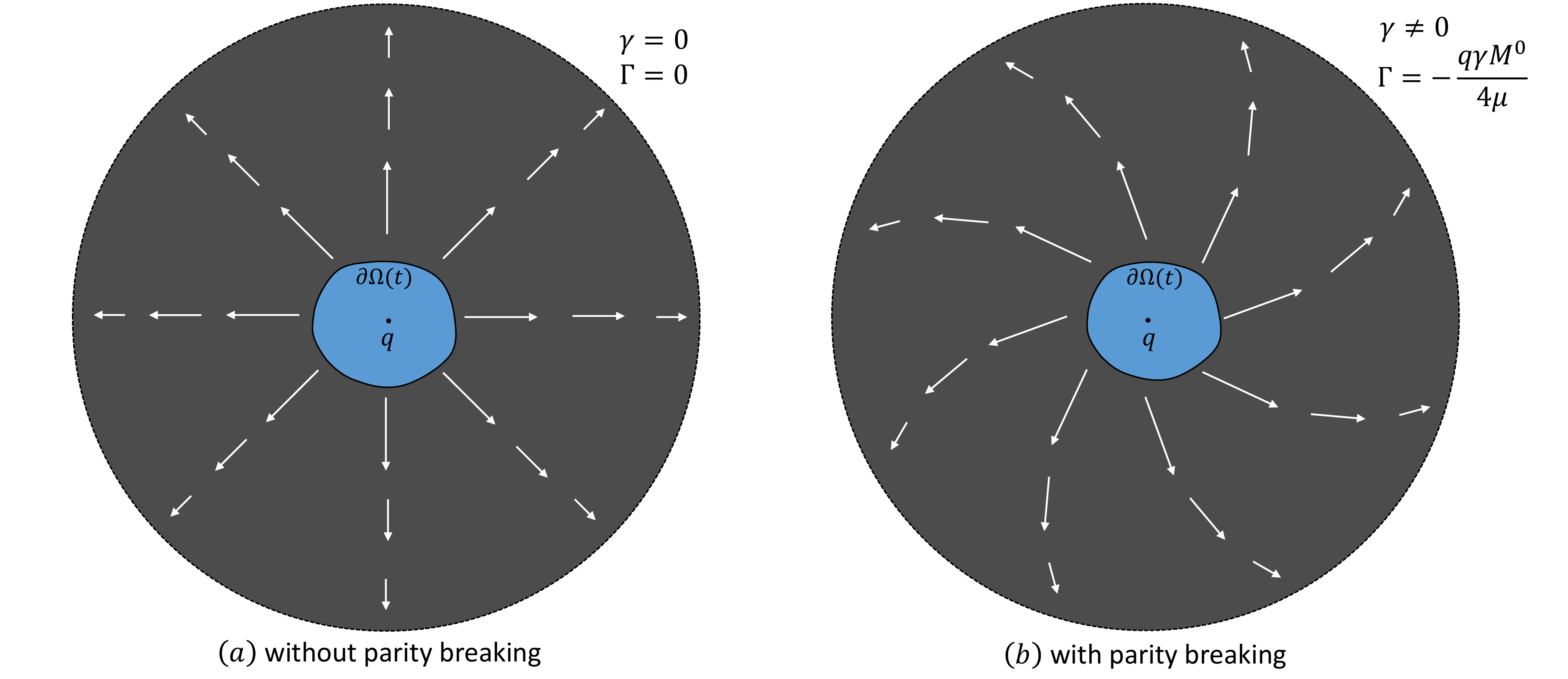}
\caption{Schematic picture of injection of a test fluid (blue) into a ferrofluid (black), with a distinct interface $\partial\Omega$ formed between them. (a) In the absence of the parity breaking coupling $\gamma$, the ferrofluid flows radially outward and has no far field circulation ($\Gamma=0$). (b) When the parity breaking coupling $\gamma$ is nonzero the ferrofluid acquires a spiral behavior everywhere in the domain. In the far field the circulation can be computed even without knowing the details of the interface.}
\label{fig:schematic}
\end{figure*}
%%%%%%%%%%%%%%%%%%%

A Hele-Shaw (HS) cell is a simplified experimental setup that could potentially test the presence of vorticity-magnetization coupling leading to parity breaking effects in ferrofluids. A HS cell consists of two parallel plates separated by an infinitesimally small gap $h$, with the hydrodynamic variables assumed to vary in the $xy$ plane at much larger length scales than the distance between the two plates. Confining ferrofluids to a HS cell has much experimental and theoretical background~\cite{langer1992labdynamics, jackson1994fingerinstability, lira2016ferroshaw, dias2015azimuthal}. A remarkable example is that of the labyrinthine instability~\cite{rosensweig1983labyrinthine}. For our purposes, we use the HS cell as a way to measure this new coupling constant $\gamma$.

Following the analysis seen in~\cite{reynolds2021oddlaw}, we take the coordinates to scale as
\begin{align}
t\sim\frac{1}{\epsilon}\,,\qquad x\sim\frac{1}{\epsilon}\,,\qquad y\sim\frac{1}{\epsilon}\bar y\,,\qquad z\sim \epsilon^0\,,
\end{align}
where $\epsilon =h/L$, with $L$ being some characteristic length in the $xy$ plane. The hydrodynamic variables scale as
\begin{align}
P\sim\frac{1}{\epsilon}\,,\quad v_x\sim \epsilon^0\,,\quad v_y\sim \epsilon^0\,,\quad v_z\sim \epsilon\,.
\end{align}
Using the relaxed form of our system, Eq. (\ref{eq:simplifiedv}), the governing equations become
\begin{align}
0 &= \partial_x v_x + \partial_y v_y + \partial_z v_z , \label{eq:tempinc} \\
\partial_x P &= \mu\partial_z^2v_x -\frac{\gamma}{4}M^0\partial_z^2 v_y , \label{eq:tempx}\\
\partial_y P &= \mu\partial_z^2v_y +\frac{\gamma}{4}M^0\partial_z^2 v_x , \label{eq:tempy}\\
\partial_z P &=0 ,
\end{align}
The solutions for velocity that satisfy the no-slip and no-penetration boundary conditions on the plates are
\begin{align}
v_x &= \frac{6z}{h^2}(h-z)V_x(x,y) ,\\
v_y &= \frac{6z}{h^2}(h-z)V_y(x,y) ,\\
v_z &= 0 ,
\end{align}
and pressure is now independent of $z$. We then substitute these solutions into (\ref{eq:tempinc}) - (\ref{eq:tempy}) and average over the plate separation, which gives
\begin{align}
0 &= \partial_x V_x + \partial_y V_y, \label{eq:2dinc}\\
-\frac{h^2}{12}\partial_x P &= \mu V_x -\frac{\gamma}{4}M^0 V_y , \label{eq:2dx}\\
-\frac{h^2}{12}\partial_y P &= \mu V_y +\frac{\gamma}{4} M^0 V_x, \label{eq:2dy}
\end{align}
The above set of equations is Darcy's law, the standard governing equation of flow within a HS cell, with an extra term coming from $M^0$. This equation is of the same form as that seen in Ref.~\onlinecite{reynolds2021oddlaw}, with $\gamma M^0/4$ playing the role of the parity odd parameter seen there.

The experimental setups discussed in~Ref.~\onlinecite{reynolds2021oddlaw} provide methods to measure this coupling. Here we will only describe a simplified setup that highlights a key feature of parity odd flows. Consider HS cell with a central injection point, where a fluid (or air) is injected with a constant rate $q$ (see Fig~\ref{fig:schematic}). Details of the inside fluid are not important for this discussion, since the following effect is independent of any intricate interface features or instabilities.

Equations~(\ref{eq:2dinc}-\ref{eq:2dy}) indicate that the ferrofluid flow is irrotational and the pressure is a harmonic function, that is,
\begin{align}
\p_x V_y-\p_y V_x&=0\,, \la{eq:2dirr}
\\
\nabla^2 P&=0\,. \la{eq:pressure}
\end{align}
It is not hard to see that $V(\zeta)\equiv V_x(x,y)-iV_y(x,y)$, with $\zeta=x+iy$, is an analytic complex function, since the Cauchy-Riemann conditions become the incompressibility and the irrotational equations~(\ref{eq:2dinc},\ref{eq:2dirr}). Assuming that the ferrofluid domain $\Omega(t)$ starts at the fluid interface and extends out to infinity, the analytic velocity function $V(\zeta)$ can be expanded in Laurent series as
\be
V(\zeta)=\sum_{n=1}^\infty\frac{c_n}{\zeta^n}\,, \la{eq:vel-anal}
\ee
where the coefficients $c_n$ are determined by the boundary conditions at the fluid interface. From the residue theorem, one can determine the coefficient $c_1$ by integrating $V(\zeta)$ over the fluid interface $\p\Omega$, i.e,
\be
2\pi i c_1=\varointctrclockwise_{\p\Omega} V(\zeta) d\zeta=\Gamma-i\frac{d\mathcal A}{dt}, \la{eq:c1}
\ee
where $\Gamma$ is the fluid circulation and $d\mathcal A/dt=-q$ is the rate of change of the area occupied by the ferrofluid.

From the expression~(\ref{eq:vel-anal}) it is possible to determine the ferrofluid pressure, which becomes
\be
P=-\frac{3}{h^2}\,\text{Re}\left[(4\mu-i\gamma M^0)\left(\frac{q-i\Gamma}{2\pi}\ln\zeta-\sum_{n=1}^\infty\frac{c_{n+1}}{n\zeta^n}\right)\right].
\ee
Since the pressure must be a single-valued function, this imposes that $(4\mu-i\gamma M^0)(q-i\Gamma)$ must be a real coefficient. This shows that the ferrofluid circulation must be proportional to
\be
\Gamma= -\frac{q\gamma M^0}{4\mu}\,.
\ee
The existence of non-zero circulation in the outer region (far field flow of ferrofluids) can be used to directly measure the coupling constant $\gamma$, and serves as a straightforward way to test for the presence or absence of the vorticity-magnetization coupling in ferrofluids. We emphasize that the far field circulation, represented by the variable $\Gamma$, is independent of the two fluid interface. The physics at the boundary between two fluids can vary greatly depending on the viscosity of the fluid being injected. For example, when a less viscous fluid such as an air bubble is injected into a ferrofluid, the Saffman-Taylor fingering instabilities can occur. These instabilities are not present when a more viscous fluid is injected. The advantage of this proposal is that the circulation in the far field, away from the interface, is not affected by the intricate and complex physics that occur near the interface.

\begin{comment}

If the central fluid is sufficiently more dense and viscous than the ferrofluid, the Saffman-Taylor instability will not be present and the interface will remain reasonably circular (see Fig~\ref{fig:schematic}).

For a constant injection rate $q = \frac{d\mathcal A}{dt}$, incompressibility implies that the inner fluid satisfies
\begin{align}
V_r^{(1)} = \frac{q}{2\pi}\frac{1}{r} ,\quad\quad V_\theta^{(1)} = 0 .
\end{align}
In the outer region, incompressibility and the kinematic boundary condition at the interface, $V_r^{(1)}\vert_{r=R}=V_r^{(2)}\vert_{r=R}$ implies
\begin{align}
V_r^{(2)} = \frac{q}{2\pi}\frac{1}{r} .
\end{align}
Crucially, since the pressure must be independent of $\theta$, the modified Darcy's law in polar form requires $0=\mu V_\theta - \frac{\gamma M}{4}V_r$. Thus, the ferrofluid aquires a velocity component in the $\theta$ direction given by
\begin{align}
V_\theta^{(2)} = \frac{q}{2\pi}\frac{\gamma M}{4\mu^{(2)}}\frac{1}{r} ,
\end{align}
where $\mu^{(2)}$ is the shear viscosity in the ferrofluid. This circulation in the outer region serves as a direct measurement of the coupling $\gamma$ and the relative importance of the vorticity-magnetization coupling. 

\end{comment}

\section{Conclusion} \la{sec:disc}

In this work we introduced a new coupling in ferrofluids between fluid vorticity and magnetization. This modification is introduced in the context of a Hamiltonian and PB framework, and after following the standard ferrofluid assumptions we arrive at a modified set of governing equations containing extra parity-odd terms. When confined to a HS cell, these extra terms manifest as off diagonal components of Darcy's Law, and provide a robust experimental setting in which to measure this new coupling.

An important conclusion from our analysis is that simply including $H_{\ell\omega}$ is not enough to capture the parity-odd effects in ferrofluids due to the small size of the particles. Additionally, the coarse-graining procedure outlined in Ref.~\onlinecite{lubensky2021anal} can be used to determine the specifics of $\gamma$ from a microscopic perspective. Future research in this direction will depend heavily on experimental confirmation of the existence of parity-breaking terms in ferrofluids.

%Additionally, the modified coefficient of $\epsilon_{ijk}\omega_j M_k$ in the equation for $M_i$ may provide insight into the reactive dynamics of magnetization, as discussed by many authors, including Feng~\cite{fang2022consistent}}. 

%Future theoretical work could examine more complex systems beyond a HS cell. For example, much work as been devoted to `spin-up' flow, where the introduction of spin viscosity, a rotational analog of shear viscosity, is introduced to explain certain experimental measurements~\cite{rinaldi2002spin, chaves2008spin}. Other works have included additional terms into the bracket (\ref{eq:gwithm}) in order to capture coupling between magnetization and linear momentum~\cite{sokolov2009hamiltonian, felderhof2011ferrobracket}. Both of these setups may be explored in the context of the vorticity-magnetization coupling, and the analysis here may provide as a framework to answer these questions.

\section{Acknowledgments} 

We thank J.C. Burton for useful discussions. This work is supported
by NSF CAREER Grant No. DMR-1944967 (SG).  DR is supported by 21st century foundation startup award from CCNY. GMM was supported by the National Science Foundation under Grant OMA-1936351.

\bibliographystyle{refstyle}
\bibliography{FerroHydro-Bibliography.bib}

\begin{thebibliography}{10}

\bibitem{yan2020hurricanes}
M.~Yan and J.~Yang.
\newblock \emph{Hurricanes on tidally locked terrestrial planets: fixed surface
  temperature experiments}.
\newblock Astronomy \& Astrophysics, \textbf{643}, A37 (2020).

\bibitem{jeong2019theOG}
D.~Jeong and F.~Schmidt.
\newblock \emph{The Odd-Parity Galaxy Bispectrum} (2019).

\bibitem{zemtsov2011rotation}
V.~Zemtsov.
\newblock \emph{Evolution of Rotation Structures in the Earth's Geological
  History} (2011).

\bibitem{avron1995viscosity}
J.~Avron, R.~Seiler, and P.~G. Zograf.
\newblock \emph{Viscosity of quantum Hall fluids}.
\newblock Physical review letters, \textbf{75}, 697 (1995).

\bibitem{avron1998odd}
J.~Avron.
\newblock \emph{Odd viscosity}.
\newblock Journal of statistical physics, \textbf{92}, 543--557 (1998).

\bibitem{tokatly2006magnetoelasticity}
I.~Tokatly.
\newblock \emph{Magnetoelasticity theory of incompressible quantum Hall
  liquids}.
\newblock Physical Review B, \textbf{73}, 205340 (2006).

\bibitem{tokatly2007new}
I.~Tokatly and G.~Vignale.
\newblock \emph{New collective mode in the fractional quantum Hall liquid}.
\newblock Physical review letters, \textbf{98}, 026805 (2007).

\bibitem{haldane2011geometrical}
F.~Haldane.
\newblock \emph{Geometrical description of the fractional quantum Hall effect}.
\newblock Physical review letters, \textbf{107}, 116801 (2011).

\bibitem{hoyos2012hall}
C.~Hoyos and D.~T. Son.
\newblock \emph{Hall viscosity and electromagnetic response}.
\newblock Physical review letters, \textbf{108}, 066805 (2012).

\bibitem{bradlyn2012kubo}
B.~Bradlyn, M.~Goldstein, and N.~Read.
\newblock \emph{Kubo formulas for viscosity: Hall viscosity, Ward identities,
  and the relation with conductivity}.
\newblock Physical Review B, \textbf{86}, 245309 (2012).

\bibitem{abanov2013effective}
A.~G. Abanov.
\newblock \emph{On the effective hydrodynamics of the fractional quantum Hall
  effect}.
\newblock Journal of Physics A: Mathematical and Theoretical, \textbf{46},
  292001 (2013).

\bibitem{hoyos2014hall}
C.~Hoyos.
\newblock \emph{Hall viscosity, topological states and effective theories}.
\newblock International Journal of Modern Physics B, \textbf{28}, 1430007
  (2014).

\bibitem{laskin2015collective}
M.~Laskin, T.~Can, and P.~Wiegmann.
\newblock \emph{Collective field theory for quantum Hall states}.
\newblock Physical Review B, \textbf{92}, 235141 (2015).

\bibitem{can2015geometry}
T.~Can, M.~Laskin, and P.~B. Wiegmann.
\newblock \emph{Geometry of quantum Hall states: Gravitational anomaly and
  transport coefficients}.
\newblock Annals of Physics, \textbf{362}, 752--794 (2015).

\bibitem{klevtsov2015geometric}
S.~Klevtsov and P.~Wiegmann.
\newblock \emph{Geometric adiabatic transport in quantum Hall states}.
\newblock Physical review letters, \textbf{115}, 086801 (2015).

\bibitem{scaffidi2017hydrodynamic}
T.~Scaffidi, N.~Nandi, B.~Schmidt, A.~P. Mackenzie, and J.~E. Moore.
\newblock \emph{Hydrodynamic electron flow and Hall viscosity}.
\newblock Physical review letters, \textbf{118}, 226601 (2017).

\bibitem{pellegrino2017nonlocal}
F.~M. Pellegrino, I.~Torre, and M.~Polini.
\newblock \emph{Nonlocal transport and the Hall viscosity of two-dimensional
  hydrodynamic electron liquids}.
\newblock Physical Review B, \textbf{96}, 195401 (2017).

\bibitem{berdyugin2019measuring}
A.~Berdyugin, S.~Xu, F.~Pellegrino, R.~K. Kumar, A.~Principi, I.~Torre, M.~B.
  Shalom, T.~Taniguchi, K.~Watanabe, I.~Grigorieva, et~al.
\newblock \emph{Measuring hall viscosity of graphene's electron fluid}.
\newblock Science, page eaau0685 (2019).

\bibitem{Monteiro2018nonresistivite}
G.~S. Denicol, X.-G. Huang, E.~Moln\'ar, G.~M. Monteiro, H.~Niemi, J.~Noronha,
  D.~H. Rischke, and Q.~Wang.
\newblock \emph{Nonresistive dissipative magnetohydrodynamics from the
  Boltzmann equation in the 14-moment approximation}.
\newblock Phys. Rev. D, \textbf{98}, 076009 (2018).

\bibitem{korving1966transverse}
J.~Korving, H.~Hulsman, H.~Knaap, and J.~Beenakker.
\newblock \emph{Transverse momentum transport in viscous flow of diatomic gases
  in a magnetic field}.
\newblock Physics Letters, \textbf{21}, 5--7 (1966).

\bibitem{banerjee2017chiral}
D.~Banerjee, A.~Souslov, A.~G. Abanov, and V.~Vitelli.
\newblock \emph{Odd viscosity in chiral active fluids}.
\newblock Nature Communications, \textbf{8} (2017).

\bibitem{lubensky2021anal}
T.~Markovich and T.~C. Lubensky.
\newblock \emph{Odd Viscosity in Active Matter: Microscopic Origin and 3D
  Effects}.
\newblock Physical Review Letters, \textbf{127}, 048001 (2021).

\bibitem{monteiro2021hamiltonian}
G.~M. Monteiro, A.~G. Abanov, and S.~Ganeshan.
\newblock \emph{Hamiltonian structure of 2D fluid dynamics with broken parity}
  (2021).

\bibitem{scheibner2020elasticity}
C.~Scheibner, A.~Souslov, D.~Banerjee, P.~Sur{\'o}wka, W.~T. Irvine, and
  V.~Vitelli.
\newblock \emph{Odd elasticity}.
\newblock Nature Physics, \textbf{16}, 475--480 (2020).

\bibitem{fruchart2022oddgas}
M.~Fruchart, M.~Han, C.~Scheibner, and V.~Vitelli.
\newblock \emph{The odd ideal gas: Hall viscosity and thermal conductivity from
  non-Hermitian kinetic theory} (2022).

\bibitem{banerjee2017odd}
D.~Banerjee, A.~Souslov, A.~G. Abanov, and V.~Vitelli.
\newblock \emph{Odd viscosity in chiral active fluids}.
\newblock Nature Communications, \textbf{8}, 1573 (2017).

\bibitem{soni2018free}
V.~Soni, E.~S. Bililign, S.~Magkiriadou, S.~Sacanna, D.~Bartolo, M.~J. Shelley,
  and W.~T. Irvine.
\newblock \emph{The odd free surface flows of a colloidal chiral fluid}.
\newblock Nature Physics, \textbf{15}, 1188--1194 (2019).

\bibitem{ganeshan2017odd}
S.~Ganeshan and A.~G. Abanov.
\newblock \emph{Odd viscosity in two-dimensional incompressible fluids}.
\newblock Physical Review Fluids, \textbf{2}, 094101 (2017).

\bibitem{abanov2018odd}
A.~Abanov, T.~Can, and S.~Ganeshan.
\newblock \emph{Odd surface waves in two-dimensional incompressible fluids}.
\newblock SciPost Physics, \textbf{5}, 010 (2018).

\bibitem{abanov2018free}
A.~G. Abanov and G.~M. Monteiro.
\newblock \emph{Free surface variational principle for an incompressible fluid
  with odd viscosity}.
\newblock Phys Rev Lett, \textbf{122}, 154501 (2019).

\bibitem{vitelli2021long}
T.~Khain, C.~Scheibner, M.~Fruchart, and V.~Vitelli.
\newblock \emph{Stokes flows in three-dimensional fluids with odd and
  parity-violating viscosities}.
\newblock Journal of Fluid Mechanics, \textbf{934}, A23 (2022).

\bibitem{reynolds2021oddlaw}
D.~Reynolds, G.~M. Monteiro, and S.~Ganeshan.
\newblock \emph{Hele-Shaw flow for parity odd three-dimensional fluids}.
\newblock Phys. Rev. Fluids, \textbf{7}, 114201 (2022).

\bibitem{rosensweig2013ferrohydrodynamics}
R.~E. Rosensweig.
\newblock \emph{Ferrohydrodynamics}.
\newblock Courier Corporation (2013).

\bibitem{rosenweig1987magneticfluids}
R.~E. Rosensweig.
\newblock \emph{Magnetic Fluids}.
\newblock Annual Review of Fluid Mechanics, \textbf{19}, 437--461 (1987).

\bibitem{neuringer1964ferrohydrodynamics}
J.~L. Neuringer and R.~E. Rosensweig.
\newblock \emph{Ferrohydrodynamics}.
\newblock The Physics of Fluids, \textbf{7}, 1927--1937 (1964).

\bibitem{shliomis1974ferro}
M.~I. Shliomis.
\newblock \emph{Magnetic fluids}.
\newblock Soviet Physics Uspekhi, \textbf{17}, 153--169 (1974).

\bibitem{shliomis1988review}
M.~Shliomis, T.~Lyubimova, and D.~Lyubimov.
\newblock \emph{Ferrohydrodynamics: An Essay On The Progress Of Ideas}.
\newblock Chemical Engineering Communications, \textbf{67}, 275--290 (1988).

\bibitem{fang2022consistent}
A.~{Fang}.
\newblock \emph{{Consistent hydrodynamics of ferrofluids}}.
\newblock Physics of Fluids, \textbf{34}, 013319 (2022).

\bibitem{muller2002structure}
H.~Müller and M.~Liu.
\newblock \emph{Structure of ferrofluid dynamics}.
\newblock Physical review. E, Statistical, nonlinear, and soft matter physics,
  \textbf{64}, 061405 (2002).

\bibitem{rosensweig1983labyrinthine}
R.~E. Rosensweig, M.~Zahn, and R.~Shumovich.
\newblock \emph{Labyrinthine instability in magnetic and dielectric fluids}.
\newblock Journal of Magnetism and Magnetic Materials, \textbf{39}, 127--132
  (1983).

\bibitem{vieu2018shape}
T.~Vieu and C.~Walter.
\newblock \emph{Shape and fission instabilities of ferrofluids in non-uniform
  magnetic fields}.
\newblock Journal of Fluid Mechanics, \textbf{840}, 455--497 (2018).

\bibitem{rinaldi2002spin}
C.~Rinaldi and M.~Zahn.
\newblock \emph{Effects of spin viscosity on ferrofluid duct flow profiles in
  alternating and rotating magnetic fields}.
\newblock Journal of Magnetism and Magnetic Materials, \textbf{252}, 172--175
  (2002).

\bibitem{chaves2008spin}
A.~Chaves, M.~Zahn, and C.~Rinaldi.
\newblock \emph{Spin-up flow of ferrofluids: Asymptotic theory and experimental
  measurements}.
\newblock Physics of Fluids, \textbf{20} (2008).

\bibitem{shliomis2021rotate}
M.~I. Shliomis.
\newblock \emph{How a rotating magnetic field causes ferrofluid to rotate}.
\newblock Phys. Rev. Fluids, \textbf{6}, 043701 (2021).

\bibitem{sokolov2009hamiltonian}
V.~V. Sokolov, V.~V. Tolmachev, and P.~A. Eminov.
\newblock \emph{Hamiltonian formalism of equations of ferrohydrodynamics}.
\newblock Doklady Physics, \textbf{54}, 488--490 (2009).

\bibitem{felderhof2011ferrobracket}
B.~U. Felderhof, V.~V. Sokolov, and P.~A. Éminov.
\newblock \emph{Hamiltonian field theory of ferrohydrodynamics}.
\newblock The Journal of Chemical Physics, \textbf{135}, 144901 (2011).

\bibitem{Note1}
This is equivalent to $\rho =1$ everywhere.

\bibitem{lakshmanan2011fascinating}
M.~Lakshmanan.
\newblock \emph{The fascinating world of the Landau--Lifshitz--Gilbert
  equation: an overview}.
\newblock Philosophical Transactions of the Royal Society A: Mathematical,
  Physical and Engineering Sciences, \textbf{369}, 1280--1300 (2011).

\bibitem{langer1992labdynamics}
S.~Langer, R.~Goldstein, and D.~Jackson.
\newblock \emph{Dynamics of labyrinthine pattern formation in magnetic fluids}.
\newblock Physical review. A, \textbf{46}, 4894--4904 (1992).

\bibitem{jackson1994fingerinstability}
D.~Jackson, R.~Goldstein, and A.~Cebers.
\newblock \emph{Hydrodynamics of fingering instabilities in dipolar fluids}.
\newblock Physical review. E, Statistical physics, plasmas, fluids, and related
  interdisciplinary topics, \textbf{50}, 298--307 (1994).

\bibitem{lira2016ferroshaw}
S.~Lira and J.~Miranda.
\newblock \emph{Ferrofluid patterns in Hele-Shaw cells: Exact, stable,
  stationary shape solutions}.
\newblock Physical review. E, Statistical physics, plasmas, fluids, and related
  interdisciplinary topics, \textbf{93} (2016).

\bibitem{dias2015azimuthal}
E.~Dias and J.~Miranda.
\newblock \emph{Azimuthal field instability in a confined ferrofluid}.
\newblock Physical Review E, \textbf{91} (2015).

\end{thebibliography}

\end{document}